\documentclass[superscriptaddress,aps,twocolumn,prb,floatfix]{revtex4}
\usepackage{eurosym}
\usepackage{amssymb}
\usepackage{amsmath}
\usepackage[dvips]{graphicx}
\usepackage{bm}

\makeatother
\begin{document}

\title[DESINT in transport through molecules with e-e and EVI interactions]{Destructive quantum interference in transport through molecules
with electron-electron and electron-vibration interactions}
\author{P. Roura-Bas}
\address{Centro At\'{o}mico Bariloche and Instituto Balseiro, Comisi\'{o}n Nacional
de Energ\'{\i}a At\'{o}mica, CONICET, 8400 Bariloche, Argentina}
\author{F. G\"uller}
\address{Dpto de F\'{\i}sica, Centro At\'{o}mico Constituyentes, Comisi\'{o}n
Nacional de Energ\'{\i}a At\'{o}mica, CONICET, Buenos Aires, Argentina}
\author{L. Tosi}
\address{Quantronics Group, SPEC, CEA, CNRS, Universit\'e Paris-Saclay, CEA
Saclay, 91191 Gif-sur-Yvette, France}
\author{A. A. Aligia}
\address{Centro At\'{o}mico Bariloche and Instituto Balseiro, Comisi\'{o}n Nacional
de Energ\'{\i}a At\'{o}mica, CONICET, 8400 Bariloche, Argentina}

\begin{abstract}
We study the transport through a molecular junction exhibiting interference effects. 
We show that these effects can still be observed in the presence of molecular vibrations if Coulomb 
repulsion is taken into account. In the Kondo regime, the conductance of the junction can be changed 
by several orders of magnitude by tuning the levels of the molecule, or displacing 
a contact between two atoms, from nearly perfect destructive 
interference to values of the order of $2e^2/h$ expected in Kondo systems. We also show that this 
large conductance change is robust for reasonable temperatures and voltages for symmetric and asymmetric tunnel 
couplings between the source-drain electrodes and the molecular orbitals. This is relevant for the 
development of quantum interference effect transistors based on molecular junctions.
\end{abstract}


\maketitle


\section{Introduction}
\label{intro}

Transport properties of single molecules are being extensively studied due
to their potential use as active components of new electronic devices \cite{naturefocus,cuevas}.
In particular, recent works \cite{carda,ke,bege,dona,rinc,benz,garner,li,bai} have focused on the possibility of using 
quantum interference effects present in molecular junctions to control the transport. 
Level (quasi-)degeneracy is very common to molecular systems
\cite{jan} providing different pathways for electron to 
tunnel, and when these levels have opposite parity, there is destructive interference
and the conductance is strongly reduced \cite{gued,ball}. 
The proposal of the {\it quantum interference effect transistor} (QuIET) \cite{carda} for example, 
is based on exploiting destructive interference (DESINT) and switching on-off the current by introducing-removing 
decoherence in a controlled manner. 

In recent experiments, the position of the highest occupied and lowest unoccupied molecular orbitals 
with respect to the Fermi level of two different molecules has been shifted using an electrochemically gated, 
mechanically controllable break junction technique \cite{li,bai}. This permitted to reduce the conductance
from values below $10^{-3} G_0$, where $G_0=2 e^2/h$ is the quantum of conductance to values of the 
order of $10^{-6} G_0$ with a small change in gating potential. More specifically, a change in voltage 
by only 17 mV caused a change in one order of magnitude in the conductance \cite{li}. 

In most of the above mentioned works, the ground state has an even number of electrons and then, the
Kondo effect is absent. In addition, the theoretical analysis often neglected electronic correlations, which 
are essential in the Kondo effect. Due to their small sizes, charging energy in molecules is very important 
\cite{yu3} and configurations with an even or 
odd number of electrons are separated by a large Coulomb energy. For an odd occupancy, 
the localized spin of the molecular orbital can be screened by the conduction electrons of the electrodes leading
to the Kondo effect.
This phenomenon is well known in condensed matter \cite{hewson} and leads to a peak in the 
conductance, at zero bias which appears at low temperatures. Signatures of the Kondo effect have been observed 
in semiconducting \cite{cro,keller} as well as in 
molecular \cite{liang,parks2,serge,park,yu2,fernandez,rak,iancu} 
quantum dots. The conductance at zero bias reaches a maximum near the quantum of conductance $G_0$, 
which in non-interacting systems can only be obtained by tuning the energy of one level near 
the Fermi energy.

In this work we explore the physics of the QuIET in the Kondo regime, where the values of 
the conductance are near $G_0$, incorporating also the effect of molecular vibrations, which are 
known to be strongly coupled to electrons, playing a major role in 
molecular junctions leading for example to finite-voltage replicas of the zero-bias peak 
of the conductance as a function of bias voltage \cite{park,yu2,fernandez,rak,iancu}.
At temperatures higher than the relevant frequencies, vibrations have been shown to strongly quench 
DESINT \cite{haer,haer2}. At small temperatures $T$ the vibrations have been argued to have a small effect \cite{ke}.
In this work we show that even at $T=0$ some effects of the vibrations remain (renormalization of the 
energies and tunnel amplitudes) 
but DESINT can be restored if the difference 
between the interference levels can be tuned.

As shown in more detail below (\ref{equi}), in the presence of two quasi-degenerate orbitals of 
opposite parity, 
it is known that the Kondo effect is enhanced due to additional degrees of freedom
(two possible orbitals and spin up/down). Interference is observed in the Kondo regime 
and the conductance can be varied from zero, total DESINT, 
to a high value $\sim G_0$ by breaking the orbital degeneracy \cite{desint}. 
Contrary to the non-interacting case, 
the levels do not need to be aligned with the Fermi level to lead to a large conductance and the effect 
relies on the formation of a many-body 
state at low temperature. However, how this picture of a  ``many-body QuIET'' 
changes for asymmetric coupling to the leads and when 
vibrations are included
is not obvious. 

In this paper, we generalize previous models describing the effect of interference on molecular transport, 
to include both Coulomb interaction which leads to the Kondo effect and electron-vibration interaction. 
Our main result is that by tuning the energy difference between the quasi-degenerate levels $\delta$ 
(by stretching the device \cite{parks2,ball} for example), DESINT can be restored 
even in the presence of vibrations. In contrast to the non-interacting case, we find that almost perfect 
DESINT persists over a wide range of voltages and temperatures for general tunnel coupling 
to the electrodes. 
We show that perfect DESINT is also expected in annulene molecules connected to the leads 
at an angle of 90 degrees (for example, one lead connected with one C atom and the other between two C atoms, 
see \ref{annu}), and changing this angle conduction is restored.
Our results show that the large tunable change in the conductance could be exploited for 
a more robust ``many-body QuIET'' operating in the Kondo regime at low bias voltage. 

For symmetric coupling to the leads, the effect can be understood as follows.
In absence of  electron-vibration interaction (EVI), as discussed in detail below, the model
has SU(4) symmetry, and the conductance is zero due to perfect DESINT. 
This symmetry is broken by the EVI permitting finite conductance.  However, in the Kondo regime
for a given $\delta$ we obtain  
the conductance drops to zero within the precision of our results and is very low for 
a scale of temperatures and voltages below a characteristic SU(4) Kondo scale. 
This is due to the fact to a very good approximation,
the SU(4) symmetry is restored as an emergent one at low energies.
For asymmetric coupling to the leads, we find numerically that tuning $\delta$ 
also the conductance falls to very small values (below $\sim 10^{-6}$), but we
cannot support this result with symmetry arguments.

The emergence of SU(4) symmetry at low energies
has been recently discussed \cite{restor,nishi} as an explanation of the observed 
SU(4) Kondo physics in a transport experiment through a double quantum dot system in spite of the different 
coupling parameters for both dots \cite{keller}. 
Here we show its relevance for the ``many-body QuIET'' in presence of EVI.

The paper is organized as follows. In Section \ref{mod} we describe the 
model and the formalism used to calculate the conductance.
Section \ref{res} contain the main results, and Section \ref{sum} contains a brief summary 
and a discussion.

\section{Model and formalism}
\label{mod}

The minimal electronic model for total DESINT consist of two 
degenerate levels of opposite parity, one coupled to both conducting leads with hybridization $V$ 
and the other with matrix elements $V$ and $-V$ respectively \cite{desint,haer} 
(see the inset of Fig. 1). 
Annulene molecules with contacts at 90 degrees are described by this model (see Refs. \cite{rinc,benz} 
and \ref{annu}).
To include the effect of vibrations in this minimal model we consider an extension to two levels of 
the Anderson-Holstein 
Hamiltonian \cite{fon,sate,corna1,paaske,hm,lili,corna2,zitko,mon,yang} in which a spin-$1/2$ doublet is 
connected to two metallic reservoirs and coupled to a phonon mode of energy $\Omega$, 
through the electron-phonon interaction $\lambda$. It is also an extension of the model proposed 
by of H\"artle \textit{et al.} \cite{haer} to include Kondo physics and general 
couplings to the leads. 

The Hamiltonian is
\begin{eqnarray}
H &=&\sum_{\sigma }\left[(E_{d}+\delta) n_{1\sigma }+ E_{d}n_{2\sigma }\right] + U\sum_{i\sigma \neq i^{\prime }\sigma ^{\prime }}n_{i\sigma }n_{i^{\prime
}\sigma ^{\prime }} \nonumber\\
&&+\sum_{\nu k\sigma }\epsilon _{k}^{\nu
}c_{\nu k\sigma }^{\dagger }c_{\nu k\sigma } +\sum_{i\nu k\sigma}(V_{ik}^{\nu }d_{i\sigma }^{\dagger }c_{\nu k\sigma }+\mathrm{H.c}.)\nonumber \\
&&+ \Omega a^{\dagger }a + \sum_{\sigma }  \lambda (a^{\dagger
}+a) n_{1\sigma},  \label{ham}
\end{eqnarray}
where $a^{\dagger }$ creates the Holstein phonon mode, $n_{i\sigma }=d_{i\sigma }^{\dagger }d_{i\sigma }$, $d_{i\sigma}^{\dagger }$ ($i=1,2)$ 
creates an electron with spin $\sigma $ at the molecular state $i$, $c_{\nu k\sigma }^{\dagger }$ creates a conduction electron at the 
left ($\nu=L$) or right ($\nu =R$) lead, and $V_{ik}^{\nu }$ describe the hopping elements between the leads and the molecular states. 
As in Ref. \cite{haer}, we only couple level 1 with the vibration (this is justified by the strong variation of the EVI depending 
on the electronic level and the vibrating mode \cite{jan}), and assume that level 1 is even and level 2 
is odd (the choice of level is irrelevant). 
We take the limit of very large Coulomb repulsion $U\rightarrow \infty $.

For $\delta=\lambda=0$, a unitary transformation (see \ref{equi}) maps the electronic part of the model into 
another one with SU(4) symmetry. This means that the Hamiltonian is invariant under a unitary change of basis 
in the 4-dimensional space of orbital and spin degrees of freedom. The conductance vanishes in this limit reflecting 
perfect DESINT \cite{desint}.

We calculate the conductance $G(V_b,T)=dI/dV_b$ by numerical differentiation of the current $I$
with respect to the bias voltage $V_b$. The expression of the current is
given by \cite{benz}
\begin{eqnarray}
I &=& \frac{ie}{h}\int d\omega {\rm Tr}[(\mathbf{\Gamma ^{L}}%
f_{L}(\omega )-\mathbf{\Gamma ^{R}}f_{R}(\omega ))\mathbf{G}_{\mathbf{d}}^{>}(\omega )  \nonumber \\
&+&(\mathbf{\Gamma ^{L}}(1-f_{L}(\omega ))-\mathbf{\Gamma ^{R}}(1-f_{R}(\omega )))\mathbf{G}_{\mathbf{d}}^{<}(\omega )],  \label{ia}
\end{eqnarray}
where the two $2\times 2$ matrices $\mathbf{\Gamma ^{\nu }}$, $\nu=L,R$ are defined by the 
matrix elements 
($\Gamma_{ij}^{\nu }= 2 \pi \sum_{k} V_{ik}^{\nu } \bar{V}_{jk}^{\nu } \delta (\omega -\epsilon_{k}^{\nu })$), 
assumed independent of $\omega$ and
$\mathbf{G}_{\mathbf{d}}^{<}$ and $\mathbf{G}_{\mathbf{d}}^{>}$  are $2\times 2$
matrices that correspond to the 
lesser $G_{ij}^{<}(\omega)$ and greater $G_{ij}^{>}(\omega)$ Green functions, 
calculated in the Keldysh formalism within 
the noncrossing approximation (NCA) \cite{win,nca2}.
We assume $\mu _{L }=eV_b/2$ and $\mu _{R}=-eV_b/2$. Our conclusions do not depend on this assumption.
This approach has been successfully applied to a variety of 
problems \cite{benz,serge,desint,restor,fon,sate,scali,tetta}, 
in particular to one magnetic level interacting with phonons \cite{fon,sate}
and two interfering magnetic 
levels without phonons \cite{benz,desint}. 
The observed scaling with temperature of the satellite peaks of the main Kondo peak due to electron 
vibration interaction \cite{rak},
is correctly reproduced by the theory \cite{sate}.
While the extension of the NCA formalism to the full problem: interference$+$Kondo$+$vibrations 
is rather straightforward, the physics of the complete system is much richer, as we show below.
Alternative treatments for non-equilibrium problems such as renormalized perturbation theory \cite{ogu,ct,tk}
are difficult to generalize for the two-level case.

We define $\Delta_{i}^{\nu }=\pi \sum_{k}|V_{ik}^{\nu }|^{2}\delta (\omega -\epsilon_{k}^{\nu })$ assumed 
independent of energy and $\Delta_{i}=\Delta_{i}^{L}+\Delta _{i}^{R}$. We take a flat band of conduction 
states extending from $-D$ to $D=1$, chosen as the unit of energy. We also fix $\Omega=0.1$, $E_d=-0.4$ and 
vary the other parameters. For $D=1$\ eV, the values of $\Omega$ and $\lambda$ that we use are 
typical \cite{haer}, while the Kondo temperature $T_K$ depends on the particular system (for example it 
varies between 1.1 
and 210\ K in Ref. \cite{parks2}).

\section{Results}
\label{res}

\subsection{Renormalization of $\delta$ by the EVI}
\label{renod}

\begin{figure}[ht]
\begin{center}
\includegraphics[clip,width=8.0cm]{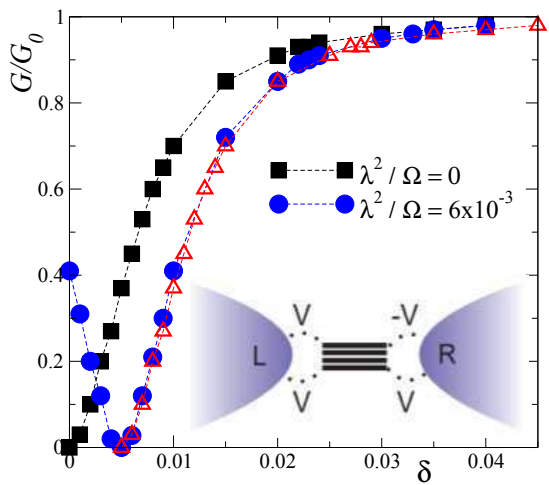}
\caption{(Color online) Differential conductance as a function of
level splitting for $\lambda=0$ and $\lambda=\sqrt{6} \times 10^{-2}$.
Other parameters are  $D=1$, $\Omega=0.1$, $\Delta_i=0.075$ and $T=0.1 T_K$, being
$T_K^{\rm SU(4)} \sim 4 \times 10^{-3}$.
Triangles correspond to
$\lambda=0$ displaced in $\delta_{\rm res}=0.005$. $G_0=2e^2/h$. (details in text below)
The inset shows a scheme of the electronic levels and their coupling to the leads.}
\end{center}
\label{1}
\end{figure}

We start by first discussing the simplest case of symmetric coupling to the leads so that $V_{1k}^{L}=V_{1k}^{R}=V_{2k}^{L}=-V_{2k}^{R}$ 
(see the inset in Fig. 1). 
Later we will change the magnitude of the hoppings but not the relative signs. 
Without vibrations ($\lambda=0$), the model for $\delta=0$, is equivalent to the SU(4) impurity Anderson model in the Kondo limit 
(see \ref{equi}). The total DESINT is reflected as a zero conductance, as shown in Fig. 1 with black squares. 
As the level detuning $\delta$ increases from zero, orbital symmetry is lost and the conductance increases. 

Much of the physics of the problem can be understood by looking at the spectral density, 
which is involved in the determination of the transport properties. The spectral density in the SU(4) case, 
$\rho _{i \sigma }(\omega )$, shows a peak near the Fermi energy (dashed-dot-dot line in the left inset 
of Fig. 2) \cite{lim,su42}. 
The half width at half maximum of this peak is of the order of the Kondo temperature for the SU(4) model 
$T_K^{\rm SU(4)}$, 
and its center is at $\omega \sim T_K^{\rm SU(4)}$, where $\omega=0$ is the Fermi energy.
For $\delta \neq 0$ the model has only spin SU(2) symmetry, but the changes in different quantities are not 
significant until $|\delta| > T_K^{\rm SU(4)}$ (Ref. \cite{desint}). 
For $\delta > T_K^{\rm SU(4)}$, the
spectral densities split and the peak in $\rho _{i \sigma }(\omega )$ for the level $i$ of lower energy narrows and displaces 
towards the Fermi level as $|\delta|$ increases, while the other is near $\omega=\delta$.
The half width at half maximum of the former peak represents the Kondo temperature $T_K$ of the system 
and lies between $T_K^{\rm SU(4)}$ for $\delta=0$ to the usual (much lower) one-level Kondo temperature 
$T_K^{\rm SU(2)}$ of the SU(2) case for $\delta \rightarrow \infty$.

The crossover SU(4)$\rightarrow$SU(2) is reflected in transport as an increase in conductance from zero 
at $\delta=0$ to a larger value as $\delta > T_K^{\rm SU(4)}$ (illustrated by squares in Fig. 1). 
For $\delta=0$, the conductance $G(V_b,T)=0$ for any bias voltage $V_b$ and  temperature $T$ due to 
DESINT \cite{desint}. At $T=0$, Fermi liquid relationships imply (see \ref{cond1})
\begin{equation}
G(0,0)=\frac{e^{2}}{h} \sum_{\sigma}
\sin ^{2} \left[ \pi (\langle n_{2 \sigma} -  n_{1 \sigma} \rangle) \right],
\label{g00}
\end{equation}
which depends on the mean occupations of the levels. For $\delta=0$, these occupations are the same and therefore, 
$G$ goes to zero. 
The dependence of $G(0,0)$ on $\delta$ is illustrated by squares in Fig. 1.
For small $\delta$, $G(0,0)>0$ but very small, and $G(0,T)$ has the temperature dependence 
characteristic of SU(4) systems \cite{desint,anders} while for $\delta > T_K^{\rm SU(4)}$, 
$G(0,T)$ approaches that of the one-level SU(2) case \cite{desint}. 

Next, we note some basic results of the Anderson-Holstein model for one 
level \cite{fon,sate,corna1,paaske,hm,lili,corna2,zitko,mon,yang}. As also showed in the experiments \cite{rak}, 
satellites of the Kondo peak appear at energies shifted from it in integers of $\Omega$. Moreover, 
the temperature dependence of the satellites follow the universal one of the main 
Kondo peak \cite{sate} 
for $\Omega \gg T_K$. The energy level is shifted by an amount $\delta_\lambda$ which for small hybridization 
is $-\lambda^2/\Omega$ (in general $-\lambda^2/\Omega < \delta_\lambda < 0$), and the effective hybridization 
is reduced \cite{fon}. This can be understood qualitatively using a 
Lang-Firsov canonical transformation \cite{haer,haer2}. 
However decoupling electrons and phonons after this transformation, predicts that the 
Kondo temperature changes exponentially with the electron-phonon coupling $\lambda$, 
which is actually not the case \cite{fon,paaske,hm,mon}.
Our NCA approach reproduces correctly the effects of the electron-vibration interaction \cite{fon,sate}.
In Fig. 1 we show the effect of this renormalization. Here we compare $G(0,T\rightarrow 0)$ 
with and without electron-vibration interaction as a function of $\delta$. 
The temperature is low enough so that the conductance is saturated to the zero-temperature value \cite{scali}. 
As stated above, for $\delta=\lambda=0$ there is perfect DESINT and the conductance vanishes. 
Keeping $\delta=0$ and turning on EVI, a sizable value of the conductance, of the order of 40 \% of the ideal one $G_0=2 e^2/h$ 
is obtained (circles in Fig. 1). Interestingly, increasing $\delta$ a situation with 
total DESINT is restored for a new value of $\delta$, which we denote as $\delta_{\rm res}$ 
(numerically $G < 3 \times 10^{-5} G_0$ for $\delta=\delta_{\rm res}=0.005$). 
This indicates that the primary effect of the EVI is just a downward correction of the energy of the corresponding 
level $\delta_\lambda \gtrsim -\lambda^2/\Omega=-0.006$ which can be compensated in the model by a positive shift 
$\delta_{\rm res}=+|\delta_\lambda|$ of the bare energy, and total DESINT is obtained again. 
This can be understood in a Fermi liquid description: at temperatures below all relevant energy scales in the model, 
phonons can be integrated out and the low-energy physics can be described by a purely electronic weakly 
interacting system with effective parameters, 
retaining the symmetry of the problem. This leads to Eq. (\ref{g00}) for the conductance at $T=0$ and changing 
$\delta$, one can render $\langle n_{2 \sigma} -  n_{1 \sigma} \rangle=0$ and then $G(0,0)=0$.
Note that for these parameters, the occupancies behave in the same way as a system with SU(4) symmetry,
although (as we shall show) small deviations with temperature and in other properties indicate
that the actual symmetry is still SU(2).
As we show below (Section \ref{nonic}), 
this phenomenon of (approximate) symmetry restoration 
is more robust than in a non-interacting system and relies on the formation of a many-body Kondo 
state at low temperatures.

\subsection{Renormalization of $\Delta_1$ and its effect on the DESINT}
\label{renoh}

\begin{figure}[ht]
\begin{center}
\includegraphics[clip,width=8.0cm]{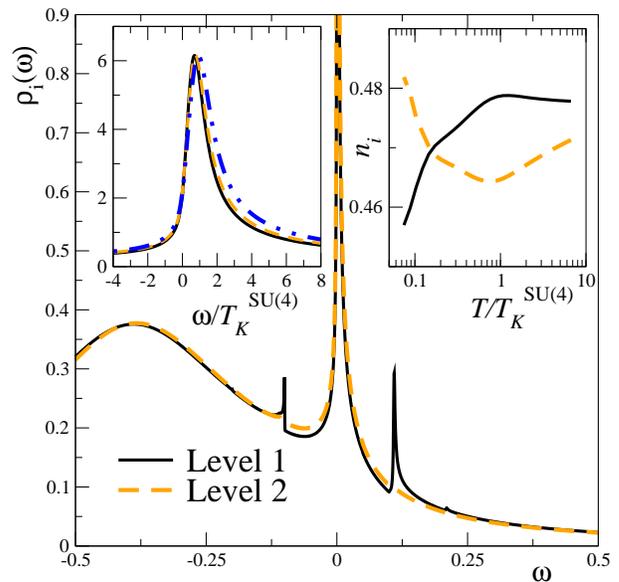}
\caption{(Color online) Spectral density of the two levels for $\lambda=\sqrt{10} \times 10^{-2}$,
$\Delta=0.05$, $\delta=\delta_{\rm res}=0.00903$ and $T=0.17 T_K^{\rm SU(4)}$ with $T_K^{\rm SU(4)}=7.5 \times 10^{-4}$. 
Other parameters as in Fig. 1. 
The left inset includes in dashed-dot-dot line the case $\lambda=\delta=0$ for comparison. 
The right inset shows the temperature dependence of the occupations $\langle n_{i \sigma} \rangle$.}
\end{center}
\label{2}
\end{figure}

However, the effect of vibrations at low temperatures is more involved than just to replace 
$\delta$ by $\delta-\delta_{\rm res}$. The reduction of the hybridization of the level coupled to the phonons, 
also plays a significant role. 
Moreover, it is not clear {\it a priori} if  perfect DESINT for some $\delta$ can persist for $T \neq 0$ 
or $V_b \neq 0$, where the Fermi liquid results cease to be valid. To show this effect more clearly, 
we have considered a larger $\lambda$ ($-\lambda^2/\Omega=-0.01$) 
and smaller $\Delta_i$ to increase the relative importance of EVI \cite{note}. In Fig. 2 
we show both spectral densities $\rho _{i \sigma }(\omega )$ 
for these new parameters and $\delta=\delta_{\rm res}=0.00903$ adjusted so that they look 
very similar near the Fermi energy. The spectral density of the level coupled to phonons 
$\rho_{1 \sigma}(\omega)$ shows satellites of the Kondo peak at $\omega \pm \Omega$ 
(as discussed for example in Ref. \cite{fon}). Small satellites at $\omega \pm 2 \Omega$ can also be noticed. 
The spectral density of level 2, $\rho_{2 \sigma }(\omega)$ 
does not have these satellites. However, strikingly, 
the Kondo peaks for both $\rho_{i \sigma}(\omega)$ are very similar and recall the corresponding result 
for SU(4) symmetry, although contrary to the Kondo peak of the SU(4) case $\delta=\lambda=0$ they are narrower 
(dashed-dot-dot line in the left inset). This can be interpreted as follows: in addition to a downward shift 
in $E_1$, the EVI causes also a downward renormalization of the magnitude of the coupling $\Delta_1$ 
(Ref. \cite{fon,khe}). 
Recent results \cite{restor,nishi} show that even if the couplings $\Delta_i$ are different, 
breaking SU(4) symmetry, this 
symmetry can be restored to a very good approximation as an emergent low-temperature symmetry by adjusting $\delta$. 
The resulting Kondo temperature corresponds to the geometrical average of both $\Delta_i$. 
In our case in which the effective $\Delta_1$ is reduced, one expects a lower Kondo temperature for the 
restored SU(4) case than to the bare one. In fact, by direct comparison with the spectral density without vibrations, 
for the parameters of Fig. 2, we find that $\Delta_1$ is reduced by a factor 0.941. 
The effects of the EVI on the DESINT can be counterbalanced by an appropriate shift in $\delta$. 
In particular, $G(0,0)=0$ by tuning $\delta$ in such a way that 
$\langle n_{2 \sigma} \rangle =  \langle n_{1 \sigma} \rangle$. 
Moreover, the occupancies have a very weak temperature dependence (see right inset of 
Fig. 2 in semi-log scale), indicating that the SU(4) symmetry restoration continues 
to be a very good approximation. The deviation between the occupancies of
both levels is below 5\% in a range of two orders of magnitude in temperature. 

\subsection{Comparison with the noninteracting case}
\label{nonic}

\begin{figure}[ht]
\begin{center}
\includegraphics[clip,width=8.0cm]{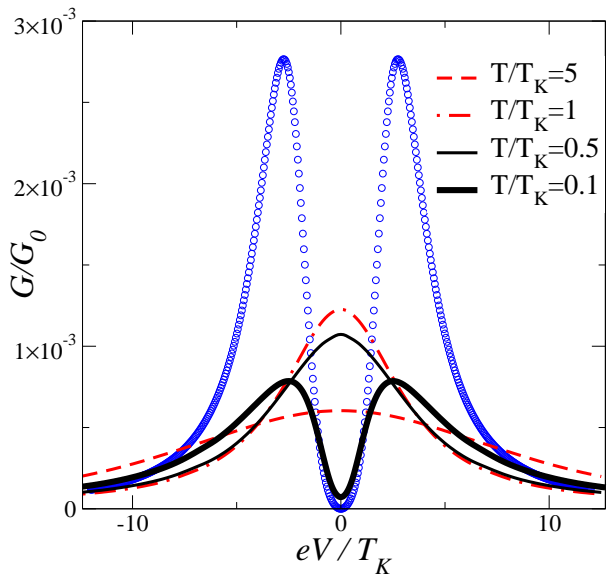}
\caption{(Color online) Conductance as a function of bias voltage for $\lambda=0$, $E_d=-0.4$
$\Delta_2=0.05$ and $\Delta_1=0.941 \Delta_2$, $\delta=0.001499$ and several temperatures.
The resulting Kondo temperature is $T_K=8 \times 10^{-4}$.
Blue circles indicate the non-interacting result at $T=0$
for $E_d=\Delta_2$ and $\delta=\Delta_1-\Delta_2$
as a function of $V_b/\Delta_2$.}
\end{center}
\label{3}
\end{figure}

The restoration of the SU(4) symmetry, and with it of the DESINT, is a
robust effect that relies on the spectral densities being almost identical
close to the Fermi level. This is very different to what can be obtained in
a non-interacting case (without Coulomb interaction and without EVI). The
appropriate way to compare both situations is the following: we consider the
interacting case for $\lambda =0$ with the same parameters of 
Fig. 2 but include the non-trivial renormalization of the hybridization 
$\Delta _{1}=0.941\Delta _{2}$. The $\delta $ is adjusted, as for Fig. 2, 
to get identical occupations of both levels at low temperatures.
The situation we want to compare is for the non-interacting case with the
same $\Delta _{i}$. In this case, for $\Delta _{1}\neq \Delta _{2}$, tuning 
\emph{both} $E_{i}$ one can also fix the two mean occupations $\langle
n_{i\sigma }\rangle =1/4$ and obtain at $V_{b}=T=0$ perfect DESINT. Since
the non-interacting spectral densities are just Lorentzian functions it
turns out that $\langle n_{i\sigma }\rangle =1/4$ implies $E_{i}=\Delta _{i}$. 
Following known equations for the non-interacting case \cite{meir}, we
obtain 

\begin{eqnarray}
G &=&\frac{e^{2}}{h}\int d\omega \left( -\frac{\partial \left[ f_{L}(\omega
)-f_{L}(\omega )\right] }{\partial \omega }\right) g(\omega ),  \nonumber \\
g(\omega ) &=&\sum\limits_{i}\frac{\Delta _{i}^{2}}{(\omega -\Delta
_{i})^{2}+\Delta _{i}^{2}}  \nonumber \\
&&-2\frac{\Delta _{1}\Delta _{2}h(\omega )}{h^{2}(\omega )+(\Delta
_{1}-\Delta _{2})^{2}\omega ^{2}},  \nonumber \\
h(\omega ) &=&(\omega -\Delta _{1})(\omega -\Delta _{2})+\Delta _{1}\Delta
_{2},  \label{gnoni}
\end{eqnarray}
where $f_{L}(\omega)=f(\omega -\mu_{L})$ and $f(\omega)$ is the Fermi function.

The transport properties for the non-interacting and Kondo cases are compared
in Fig. 3. The conductance at low temperatures in the interacting
case (black thick) and the non-interacting case (blue circles) is shown as a
function of the bias voltage scaled with the relevant scale in each case:
the Kondo temperature $T_{K}$ in the former and the hybridization $\Delta
_{2}$ in the latter. In the non-interacting case, with this tuning of both
energies, we realize a situation where the device can be operated as a
QuIET: changing $E_{2}-E_{1}$ by a quantity larger than $\Delta _{i}$, a
conductance of the order of $G_{0}$ can be reached. However, since the
spectral densities are different, there is no emergent symmetry, perfect
DESINT is rapidly lost for small $V_{b}\sim \Delta _{2}$ as shown in Fig. 2. 
Instead, in the interacting Kondo case, the regime to operate
the \textquotedblleft many-body QuIET \textquotedblright\ is found tuning
just \emph{one} energy $E_{i}$ and perfect DESINT is obtained with a total
occupancy near 1, more robust under $V_{b}$. The conductance remains small
even for $V_{b}\sim T_{K}$ and $T\sim 5T_{K}$. This is expected because the
spectral densities of both levels are very similar, as shown in Fig. 2.

\subsection{Effect of $\delta$ on $T_K$}
\label{tkd}

\begin{figure}[ht]
\begin{center}
\includegraphics[clip,width=7.5cm]{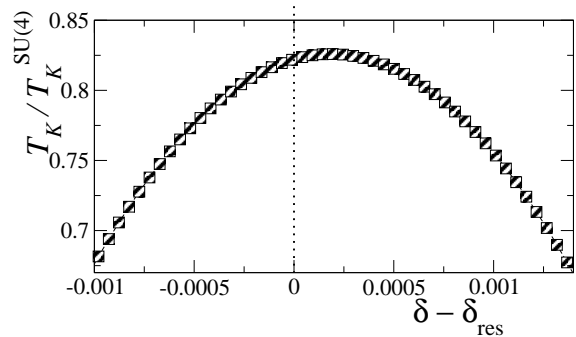}
\caption{Kondo temperature as a function of $\delta$. Other parameters as in Fig. 2 .}
\end{center}
\label{4}
\end{figure}

To discuss in more detail the width of the Kondo peak (which is of the order of  $2T_K$), 
we represent
$T_K/T_K^{\rm SU(4)}$ as a function of $\delta$ in Fig. 4. $T_K^{\rm SU(4)}$ is the Kondo temperature
of the pure SU(4) case $\delta=\lambda=0$. For practical purposes, we have calculated $T_K$ 
from the temperature dependence of the conductance $G_e(T)$ of an equivalent model for an SU(4) 
system under a symmetry breaking field. Specifically $T_K$ is determined from $G_e(T_K)=G_M/2$, where 
$G_M$ is $G_e(0)$ for a total occupation of one electron in the dot ($G_M$ is the maximum possible value of
the conductance that can be obtained in the model for fixed coupling to the leads, 
changing the other parameters, see \ref{cond2}).
This definition is more precise than the width of the spectral density or the peak in $G(V_b)$ because they 
depend on details of the fit and the proximity of the charge-transfer peak \cite{tk}.
We have checked that the ratio $T_K/T_K^{\rm SU(4)}$ calculated in this way or from the width at half maximum of the 
Kondo peak are practically the same. The maximum $T_K$ is about 17 \% below $T_K^{\rm SU(4)}$ due to EVI.
Note that this maximum is reached for a value $\delta=\delta_m$ slightly larger than $\delta_{\rm res}$,
indicating again that the emergent SU(4) symmetry is approximate and depends on the property studied.
As $\delta$ deviates from $\delta_m$, $T_K$ decreases, and more strongly for $|\delta-\delta_m|>T_K^{\rm SU(4)}$.

\subsection{Asymmetric case}
\label{asym}

\begin{figure}[h]
\begin{center}
\includegraphics[clip,width=8.0cm]{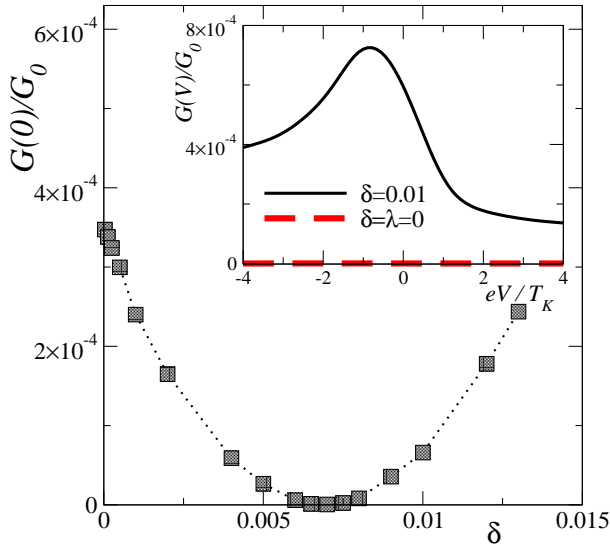}
\caption{(Color online) Zero-bias differential conductance as a function of
level splitting for strongly different lead
couplings $\Delta_i^L = 0.075$, $\Delta_i^R=0.0025$.
Other parameters are  $\Omega=0.1$, $\lambda=\sqrt{10} \times 10^{-2}$
and $T=T_K^{\rm SU(4)}/20$ with $T_K^{\rm SU(4)} \sim 6 \times 10^{-3}$.
The inset shows the differential conductance as a function of bias voltage for
$\lambda=0$ and two values of $\delta$.}
\end{center}
\label{5}
\end{figure}

Having demonstrated that perfect DESINT, destroyed after including vibrations, can be restored tuning $\delta$ for symmetric 
leads ($V_{1k}^{L}=V_{1k}^{R}$ and $V_{2k}^{L}=-V_{2k}^{R}$), one would like to know to what extent this is true for strongly asymmetric leads. 
A situation that is commonly found in molecular junctions. From the transformations shown in \ref{conda}, 
we know that $G(0,0)=0$ if $\lambda=\delta=0$ and $V_{1k}^{L}=V_{2k}^{L} > V_{1k}^{R}= - V_{2k}^{R} > 0$. 
This model describes in particular annulene molecules (such as 18-annulene \cite{ke}) with contacts at 90 degrees, 
one of then in contact with a C atom and the other in between two C atoms (see \ref{annu}). 
This perfect DESINT is lost for $\lambda>0$. 

To study the effect of a finite $\lambda$ we have taken $\Delta_i^L = 30 \Delta_i^R$, 
keeping the ratios of $V_{ik}^{\nu}$ as above. 
This large value is chosen to have a great contrast to the symmetric case. In addition, 
large asymmetry of the coupling is the general case in molecular quantum dots \cite{serge}.
In Fig. 5 we show a similar comparison to that of 
Fig. 1 of the conductance calculated with and without phonons in this asymmetric situation. 
As shown in the inset, for $\lambda=0$ the conductance $G=0$ for $V_b=0$ at low temperatures in agreement 
with our analytic calculations (\ref{conda}). As shown by the numeric results. this continues to be true for 
$V_b \ne 0$. In the inset we also show $G(V_b)$ for $\delta=0.01$. This curve shows a marked asymmetric behavior 
and a peak for $eV_b=-T_K$. 

This can be understood as follows. For an asymmetry greater than a factor 
10 between left and right couplings, $G(V_b)$ and the spectral densities
$\rho _{i \sigma }(\omega )$ are quite similar to those for the 
dot considered at equilibrium with the lead with larger 
coupling \cite{tk} (left in our case). The conductance is then similar to that in scanning tunneling spectroscopy 
experiments, with the right lead playing the role of the tip with the difference that 
in our case half of the (instead of the whole) voltage falls between the molecule and the right lead.
giving rise to capacitance effects \cite{capac}. 
For $eV_b=-T_K$, the potential of the left
lead is at $\mu_L=-T_K/2$. For the SU(4) Kondo effect, the Kondo peak in $\rho _{i \sigma }(\omega )$ 
is $\sim T_K$ above the Fermi level (see the inset in Fig. 2). Then for this nonequilibrium
case the peak lies at $\mu_L + T_K =T_K/2$, which is precisely $\mu_R$, leading to a maximum in $G(V_b)$.

For other values of $\delta$ and $\lambda$, the shape of $G(V_b)$ is practically the same, 
but the overall magnitude changes reflecting the different degree of DESINT. In particular, for 
$\lambda=\sqrt{10} \times 10^{-2}$ and $\delta=0.007$, we find $G(V_b)< 10^{-6} G_0$ for $|eV_b| < 4 T_K$.  

As expected, the perfect DESINT leading to $G=0$ for $\delta=V_b=T=0$ is destroyed as the EVI is turned 
on (black squares for $\lambda\neq 0$). However, consistent with our previous findings, 
tuning $\delta=\delta_{\rm res}$ the conductance can be made to vanish within numerical precision, 
although in this case we could not prove analytically that $G(0,0)=0$. For $\lambda$ such 
that $\lambda^2/\Omega=0.01$, $\delta_{\rm res}=0.007$ and we find $G(V_b)< 10^{-6} G_0$ for $|eV_b| < 4 T_K$.  
When $\delta$ is changed by values higher than $T_K$ the conductance reaches high values which for the 
asymmetry ratio chosen (30/1) are of the order of $0.125 G_0$.
This should be compared with values below $2 \times 10^{-6}$ that we obtain for the conductance 
in the range $- 5 T_K < V_b < 5 T_K$ and $T < T_K$.

\section{Summary and discussion}
\label{sum}
In summary, we have calculated the conductance through a molecular junction showing interference effects 
in the presence of 
electron-vibration and strong electron-electron interactions, in the Kondo regime. 
We have shown that for temperatures and voltages such that $k_B T$ and $e V_b$ are below the relevant vibration 
energies (in particular below the lowest one with important electron-phonon interaction), 
the vibrations have two effects: (i) a renormalization of the energy level most strongly coupled to the mode 
and (ii) a non-trivial renormalization of the coupling to the leads. 
These effects break the SU(4) symmetry of the simplest interference model and therefore, they are expected 
to cancel the effects of destructive interference and lead to a sizable conductance. This is the case in the 
non-interacting system. However in the interacting case for an odd occupancy of the molecule (Kondo regime), 
and symmetric coupling to the leads, 
tuning the difference between the interacting levels $\delta$, the destructive interference can be restored 
to a large extent due to a 
subtle many-body effect leading to an emergent SU(4) symmetry.
For large asymmetric coupling to the leads we find a similar result for the destructive interference,
although in this case, the result seems to be not directly related with approximate SU(4) symmetry. 

We obtain that for temperature or voltage ranging from 0 to a few times the Kondo temperature $T_K$, 
destructive interference is almost perfect for a fixed value of the energy difference between the levels $\delta$. 
Changing $\delta$ by an amount larger than $T_K$ the conductance increases by more than three orders of magnitude reaching 
values of the order of the quantum of conductance $2e^2/h$. Our results show the robustness of the ``many-body QuIET'' which 
works in the Kondo regime (for values of the on-site energies $-E_d \gg \Delta$). To obtain similar large changes in the 
conductance in a non-interacting system it is necessary that the system is in the intermediate valence regime $|E_d| \sim \Delta$. 
Moreover for a non-interacting system the destructive interference is more fragile under application of small 
bias voltages and temperatures due to the lack of emergent SU(4) symmetry. 

The presence of the Kondo effect and nearly degenerate levels of opposite parity is very common in molecular systems.
If the difference between the energy of these levels $\delta$ can be controlled,  the corresponding two-level 
Kondo effect that we have described seems promising for 
a quantum interference effect transistor, operating at low bias voltage and hence with low power dissipation.
In fact mechanically controllable break junctions have shown a great tunability of several parameters, like 
the relative position of the Fermi level \cite{li,bai} and a Kondo temperature ranging from 
near 1 to 200 K \cite{parks2}.
It is natural to expect that stretching the device, as done for example in Ref. \cite{parks2}, $\delta$
can also be varied.
Another possibility is to control the angle at which the leads are connected to an annulene molecule, 
as described in \ref{annu}.

\section*{Acknowledgments}
We thank Bruce Normand for useful discussions. We acknowledge financial support provided by PIP 112-201501-00506 
of CONICET and and PICT 2017-2726 of
ANPCyT.

\appendix
\section{Equivalence of the interference model and transport through an
SU(4) QD under a magnetic field}

\label{equi}

The model for destructive interference (DESINT) is like Eq. (\ref{ham}) of the main text
without phonons [last line of Eq. (\ref{ham})]:

\begin{eqnarray}
&&H=\sum_{i\sigma }E_{d}n_{i\sigma }+\delta \sum_{\sigma }n_{1\sigma
}+\sum_{\nu k\sigma }\epsilon _{k}^{\nu }c_{\nu k\sigma }^{\dagger }c_{\nu
k\sigma } \nonumber \\
&&+\sum_{i\nu k\sigma }(V_{ik}^{\nu }d_{i\sigma }^{\dagger }c_{\nu
k\sigma }+\mathrm{H.c}.)+U\sum_{i\sigma \neq i^{\prime }\sigma ^{\prime
}}n_{i\sigma }n_{i^{\prime }\sigma ^{\prime }}  \label{hinter}
\end{eqnarray}%

We assume identical left and right leads with equal coupling to the two levels, 
and one symmetric and one antisymmetric molecular level with splitting $\delta $. 
Specifically 
$V_{1k}^{L}=V_{1k}^{R}$, and $V_{2k}^{L}=-V_{2k}^{R}$. A schematic
representation of the model is in Fig. 6.

\begin{figure}[h]
\begin{center}
\includegraphics[clip,width=8.0cm]{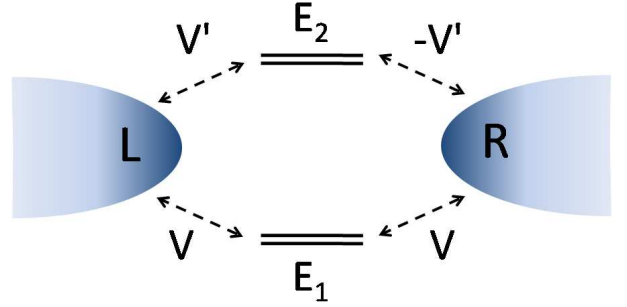}
\caption{(Color online) Scheme of the Hamiltonian Eq. (\protect\ref{hinter})}
\end{center}
\label{6}
\end{figure}

Transport through a nanotube quantum dot (QD) \cite{anders} or a Si QD in a
Si fin-type field effect transistors \cite{tetta} is characterized by a
valley (or pseudospin) degree of freedom in addition to the spin one. The
spin and pseudospin degrees of freedom can be interchanged, so that the
Zeeman splitting can be replaced by a pseudospin splitting. The model that
describes the system differs from the above one not only in the symmetry of
the levels, but also in the presence of the pseudospin index in the leads:

\begin{eqnarray}
&&H=\sum_{i\sigma }E_{d}n_{i\sigma }+\delta \sum_{\sigma }n_{1\sigma
}+\sum_{\nu ik\sigma }\epsilon _{k}^{\nu }c_{\nu ik\sigma }^{\dagger }c_{\nu
ik\sigma } \nonumber\\
&&+\sum_{i\nu k\sigma }(V_{ik}^{\nu }d_{i\sigma }^{\dagger }c_{\nu
ik\sigma }+\mathrm{H.c}.)+U\sum_{i\sigma \neq i^{\prime }\sigma ^{\prime
}}n_{i\sigma }n_{i^{\prime }\sigma ^{\prime }}.  \label{ham2}
\end{eqnarray}%
By symmetry, the matrix elements $V_{ik}^{\nu }$ should be independent of
pseudospin index, but we allow for different matrix elements for the two
levels. The same model describes a system with 
two dots capacitively coupled.\cite{keller} This model is represented in Fig. 7.

\begin{figure}[h]
\begin{center}
\includegraphics[clip,width=8.0cm]{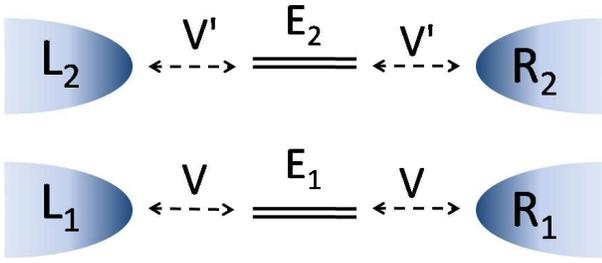}
\caption{(Color online) Scheme of the Hamiltonian Eq. (\ref{ham2})}
\end{center}
\label{7}
\end{figure}

While there are four spin degenerate bands of mobile electrons in the leads,
depending on valley index $i$ or position with respect to the quantum dot
(left or right), for each energy $\epsilon _{k}^{L}=\epsilon _{k^{\prime
}}^{R}$ for which there are states at the left and the right leads, only the
linear combination $f_{ik\sigma }^{\dagger }=(V_{ik}^{L}c_{Lik\sigma
}^{\dagger }+V_{ik^{\prime }}^{R}c_{Rik^{\prime }i\sigma }^{\dagger })/%
\tilde{V}_{ik}$ hybridizes with the state $d_{i\sigma }^{\dagger }$. while
the orthogonal one decouples. The normalization factor is
\begin{equation}
\tilde{V}_{ik}=[(V_{ik}^{L})^{2}+(V_{ik}^{R})^{2}]^{1/2}.
\label{vtil}
\end{equation}
Then, the model Eq. (\ref{ham2}) reduces to

\begin{eqnarray}
&&H=\sum_{i\sigma }E_{d}n_{i\sigma +\delta} \sum_{\sigma }n_{1\sigma
}+\sum_{ik\sigma }\epsilon _{k}^{L}f_{ik\sigma }^{\dagger }f_{ik\sigma
} \nonumber \\
&&+\sum_{i\nu k\sigma }(\tilde{V}_{ik}d_{i\sigma }^{\dagger }f_{ik\sigma }+%
\mathrm{H.c}.)+U\sum_{i\sigma \neq i^{\prime }\sigma ^{\prime }}n_{i\sigma
}n_{i^{\prime }\sigma ^{\prime }}.  \label{h3}
\end{eqnarray}

When $\delta =0$ and $\tilde{V}_{ik}$ is independent of $i$, the model
reduces to the SU(4)\ impurity Anderson model.

For the model Eq. (\ref{hinter}) we can define

\begin{eqnarray}
f_{1k\sigma }^{\dagger } &=&\frac{1}{\sqrt{2}}\left( c_{Lik\sigma }^{\dagger
}+c_{Rik\sigma }^{\dagger }\right) ,  \nonumber \\
f_{2k\sigma }^{\dagger } &=&\frac{1}{\sqrt{2}}\left( - c_{Lik\sigma }^{\dagger
} + c_{Rik\sigma }^{\dagger }\right) ,  \label{basis}
\end{eqnarray}%
and the model tales the same form as Eq. (\ref{h3}) with $\tilde{V}_{ik}=%
\sqrt{2}V_{ik}^{L}$.

\section{Conductance for the SU(4) QD with symmetry reduced to SU(2)}

\label{cond2}

For the model Eq. (\ref{ham2}) it is reasonable to assume couplings 
$\Delta_{i}^{\nu }=\pi \sum_{k}|V_{ik}^{\nu }|^{2}\delta (\omega
-\epsilon_{k}^{\nu })$ independent of energy. Then, the total current is
given by the sum of four independent contributions for each spin and
pseudospin $I=\sum_{i\sigma }I_{i\sigma }$. 

Assuming also
\begin{equation}
V_{1k}^{L }/V_{1k}^{R }=V_{2k}^{L }/V_{2k}^{R },
\label{propo}
\end{equation}
which implies $\tilde{V}_{ik}$ independent of $i$ [see Eq. (\ref{vtil})] and therefore  
SU(4) symmetry for $\delta =0$,
for each of the four channels
one can use the expression of Meir and Wingreen,\cite{meir} leading to

\begin{equation}
I_{i\sigma }(V)=\frac{\pi e}{h}\frac{\Delta_{i}^{L}\Delta _{i}^{R}} {%
\Delta_{i}}\int d\omega \rho _{i \sigma }(\omega )\left( f(\omega -\mu
_{L})-f(\omega -\mu _{R})\right) ,  \label{current}
\end{equation}
where $\Delta_{i}=\Delta_{i}^{L}+\Delta _{i}^{R}$, $\rho _{i\sigma }(\omega )
$ is the spectral density of states of level $i$ with spin $\sigma $, $%
f(\omega )$ is the Fermi distribution and $\mu _{L}-\mu _{R}=eV_b$, where $V_b$
is the applied bias voltage. For $V_b\rightarrow 0 $ the contribution to the
conductance $G_{i\sigma }(V_b)=dI_{i\sigma }(V_b)/dV_b$ is

\begin{equation}
G_{i\sigma }(0)=\frac{\pi e}{h}\frac{\Delta _{i}^{L}\Delta _{i}^{R}}{%
\Delta_{i}} \int d\omega \left( -\frac{\partial f(\omega )}{\partial \omega }%
\right) \rho _{i\sigma }(\omega ).  \label{gi}
\end{equation}

At zero temperature, it is known that the system is a Fermi liquid and since
spin and pseudospin are conserved, $\rho _{i\sigma }(\omega )$ can be
related to the phase shift $\delta _{i\sigma }$ (Ref. \cite{yoshi,lang}). For
constant density of states of the leads, the generalized Friedel sum rule,
for the case in which SU(4) symmetry is reduced only to SU(2),\cite{yoshi}
gives

\begin{equation}
\rho _{i\sigma }(\omega )= \frac{1}{\pi \Delta_i} \sin ^{2}(\delta_i) ,
\label{rho}
\end{equation}
where

\begin{equation}
\delta _{i\sigma }= \pi \langle n_{i \sigma} \rangle,  \label{delta}
\end{equation}%
so that at $T=0$

\begin{eqnarray}
G(0) &=&\sum_{i\sigma }G_{i\sigma }(0)  \nonumber \\
G_{i\sigma }(0) &=&\frac{e^{2}}{h}\sin ^{2}(\pi \langle n_{i\sigma }\rangle )
\label{g0}
\end{eqnarray}

In the Kondo limit for $U\rightarrow \infty$, one has in the SU(4) case $%
\langle n_{i \sigma}\rangle \rightarrow 1/4$, and therefore $G(0)=2e^2/h$.
We define the Kondo temperature $T_K$ as the temperature at which the
conductance $G(V_b,T)$ at $V_b=0$ falls to half of the $T=0$ value, i. e.

\begin{equation}
G(0,T_K)=e^2/h  \label{T_K}
\end{equation}

The resulting value of $T_{K}$ is similar (and in general more reliable \cite{tk}) 
to other possible definitions of $T_{K}$, 
such as the width of the peak nearer to the Fermi
energy in the total spectral density,\cite{desint} as we have verified
explicitely in our case. Since the models Eq. (\ref{hinter}) and Eq. (\ref{ham2}) are equivalent, 
we can use this definition of $T_{K}$ also for
DESINT in spite of the fact that the conductance for the
model Eq. (\ref{hinter}) vanishes for some parameters (see next section).

\section{Conductance at $T=0$ for the interference problem}

\label{cond1}

While for arbitrary temperatures, a full non-equilibrium
calculation is necessary to compute the conductance of the interference model,
Eq. (\ref{hinter}), at $T=V_b=0$ we can derive the conductance
using
the Fermi liquid properties discussed in \ref{cond2} for the equivalent problem Eq. 
(\ref{h3}).

We know the scattering matrix $S^{f\sigma }$ in the basis of the
states corresponding to the $f_{ik\sigma }^{\dagger }$ \cite{datta,pust}

\begin{equation}
S^{f\sigma }=\left( 
\begin{array}{cc}
\exp (2i\delta _{1\sigma }) & 0 \\ 
0 & \exp (2i\delta _{2\sigma })%
\end{array}%
\right) ,  \label{sf}
\end{equation}%
and we also know from Eq. (\ref{basis}) the transformation matrix to left
and right leads

\begin{equation}
U=\frac{1}{\sqrt{2}}\left( 
\begin{array}{cc}
1  & 1 \\ 
-1 & 1
\end{array}%
\right) .  \label{um}
\end{equation}
so that 

\begin{equation}
S^{c\sigma }=US^{f\sigma }U^{\dagger }.  \label{slr}
\end{equation}

The conductance at $T=V_b=0$ is given by the off-diagonal element of $S^{c\sigma }$ \cite{datta,pust}

\begin{equation}
G=\frac{e^{2}}{h}\sum_{i\sigma }|S_{LR}^{c\sigma }|^{2},  \label{g02}
\end{equation}%
which using the above expressions and some algebra takes the form

\begin{equation}
G=\frac{e^{2}}{h}\sum_{\sigma }\sin ^{2}\left( \delta _{2\sigma }-\delta
_{1\sigma }\right) ,  \label{gf}
\end{equation}%
and $\delta _{i\sigma }$ are given by Eq. (\ref{delta}).

\subsection{Case $\delta =0$ and arbitrary constant couplings}
\label{conda}

Here we generalize the above result for $\delta =0$ and $V_{ik}^{\nu}=V_{i}^{\nu }$ 
independent of $k$ but otherwise arbitrary. The independence
of $k$ is usually well justified because the range of energies that
determine $T_{K}$ is much smaller than that of the variation of  $V_{ik}^{\nu }$. 
The model is appropriate for annulene molecules in which the relevant configuration with
odd number of particles has orbital degeneracy, with two states of opposite angular momenta.\cite{rinc} 
One example is benzene.\cite{benz}
The hybridization term in  Eq. (\ref{hinter}) is

\begin{equation}
H_{V}=\sum_{k\sigma }H_{k\sigma }{\rm , } H_{k\sigma }=\sum_{i\nu
}(V_{i}^{\nu }d_{i\sigma }^{\dagger }c_{\nu k\sigma }+\mathrm{H.c}.).
\label{hv}
\end{equation}%
Then 

\begin{eqnarray}
&&\left[ H_{k\sigma },d_{i\sigma }^{\dagger }\right] 
= \sum_{\nu }\bar{V}_{i}^{\nu }c_{\nu ik\sigma }^{\dagger }, \nonumber \\
&&\left[ H_{k\sigma }, \left[H_{k\sigma },d_{i\sigma }^{\dagger }\right] \right]  
=\sum_{j\nu }\bar{V}_{i}^{\nu }V_{j}^{\nu }d_{j\sigma }^{\dagger }.  \label{com}
\end{eqnarray}
Now we look for new operators $\tilde{d}_{l\sigma }^{\dagger }$, linear
combinations of $d_{1\sigma }^{\dagger }$ and $d_{2\sigma }^{\dagger }$ such
that

\begin{equation}
\left[ H_{k\sigma },\left[ H_{k\sigma },\tilde{d}_{l\sigma
}^{\dagger }\right] \right] =\Lambda \tilde{d}_{l\sigma }^{\dagger }.
\label{hh}
\end{equation}%
Clearly, the problem is equivalent to diagonalize the quadratic form 
$\sum_{ij\nu }\bar{V}_{i}^{\nu }V_{j}^{\nu }d_{j\sigma }^{\dagger }d_{i\sigma
}$. The corresponding matrix is Hermitian, so that it can always be
diagonalized. Furthermore, it can be easily checked that the matrix is
positive definite, so that $\Lambda \geq 0$. We denote the two eigenvalues
as  $(\tilde{V}_{l})^{2}$ with  $\tilde{V}_{l}\geq 0$. Clearly the operators
that result from the diagonalization satisfy canonical anticommutation rules 
$\{\tilde{d}_{i\sigma }^{\dagger },\tilde{d}_{j\sigma }\}=\delta _{ij}$.

We now define new conduction operators by

\begin{equation}
\left[ H_{k\sigma },\tilde{d}_{l\sigma }^{\dagger }\right] =\tilde{V}%
_{l}f_{lk\sigma }^{\dagger }.  \label{fnew}
\end{equation}%
If for one $l$, $\tilde{V}_{l}=0$, the corresponding $f_{lk\sigma }$ is
defined by orthogonality $\{f_{1k\sigma }^{\dagger },f_{2k\sigma }\}=0$ and
normalization $\{f_{lk\sigma }^{\dagger },f_{lk\sigma }\}=1$.  In general
one has

\begin{eqnarray}
&&\tilde{V}_{i}\tilde{V}_{j}\{f_{ik\sigma }^{\dagger },f_{jk\sigma }\}
=\left\{ \left[ H_{k\sigma },\tilde{d}_{i\sigma }^{\dagger }\right] ,\left[
\tilde{d}_{j\sigma },H_{k\sigma }\right] \right\}   \nonumber \\
&=&\left[ H_{k\sigma },\tilde{d}_{i\sigma }^{\dagger }\right] 
(\tilde{d}_{j\sigma }H_{k\sigma }-H_{k\sigma }\tilde{d}_{j\sigma }) \nonumber \\
&&+(\tilde{d}%
_{j\sigma }H_{k\sigma }-H_{k\sigma }\tilde{d}_{j\sigma })\left[ H_{k\sigma },%
\tilde{d}_{i\sigma }^{\dagger }\right]   \nonumber \\
&&=-\tilde{d}_{j\sigma }\left[ H_{k\sigma },\tilde{d}_{i\sigma }^{\dagger }%
\right] H_{k\sigma }-\left[ H_{k\sigma },\tilde{d}_{i\sigma }^{\dagger }%
\right] H_{k\sigma }\tilde{d}_{j\sigma } \nonumber \\
&&+\tilde{d}_{j\sigma }H_{k\sigma } 
\left[ H_{k\sigma },\tilde{d}_{i\sigma }^{\dagger }\right] +H_{k\sigma }%
\left[ H_{k\sigma },\tilde{d}_{i\sigma }^{\dagger }\right] \tilde{d}%
_{j\sigma }  \nonumber \\
&&=\tilde{d}_{j\sigma }\left[ H_{k\sigma },\left[ H_{k\sigma },
\tilde{d}_{i\sigma }^{\dagger }\right] \right] 
+\left[ H_{k\sigma },\left[ H_{k\sigma
},\tilde{d}_{i\sigma }^{\dagger }\right] \right] \tilde{d}_{j\sigma } \nonumber \\
&&=(\tilde{V}_{i})^{2}\{\tilde{d}_{i\sigma }^{\dagger },\tilde{d}_{j\sigma }\}=(%
\tilde{V}_{i})^{2}\delta _{ij}.  \label{anti}
\end{eqnarray}%
Then if both $\tilde{V}_{l}>0$, $\{f_{ik\sigma }^{\dagger },f_{jk\sigma
}\}=\delta _{ij}$ automatically.

Clearly in the new basis 

\begin{equation}
H_{k\sigma }=\sum_{l}(\tilde{V}_{l}\tilde{d}_{l\sigma }^{\dagger
}f_{lk\sigma }+\mathrm{H.c}.).  \label{vnew}
\end{equation}%
Then, for $\delta =0$, the Hamiltonian takes the same simple form as Eq. (\ref{h3})
but with different localized operators ($\tilde{d}_{l\sigma }^{\dagger }$
instead of $d_{i\sigma }^{\dagger }$). 
This transformation can greatly simplify the treatment of annulene molecules at arbitrary temperatures.
In particular, in the calculation of the conductance \cite{benz} off-diagonal elements and complex numbers disappear. 

The unitary matrix that changes the
basis from the $f_{lk\sigma }^{\dagger }$ to the $c_{\nu k\sigma }^{\dagger }$ has the
general form 

\begin{equation}
U=\left( 
\begin{array}{cc}
\alpha  & \beta  \\ 
-\bar{\beta} & \bar{\alpha}%
\end{array}%
\right) .  \label{um2}
\end{equation}%
Then, using Eqs. (\ref{delta}), (\ref{sf}), (\ref{slr}), and (\ref{g02}) we
obtain

\begin{equation}
G=4|\alpha \beta |^{2}\frac{e^{2}}{h}\sum_{\sigma }\sin ^{2}\left[ \pi 
\left(\langle \tilde{n}_{2\sigma }\rangle -\langle \tilde{n}_{1\sigma }\rangle
\right) \right] ,  \label{gf2}
\end{equation}
where  the occupancies $\tilde{n}_{l\sigma }=\tilde{d}_{l\sigma }^{\dagger }
\tilde{d}_{l\sigma }$ correspond to the states that diagonalize $\sum_{ij\nu
}\bar{V}_{i}^{\nu }V_{j}^{\nu }d_{j\sigma }^{\dagger }d_{i\sigma }$.

\section{Kondo effect and destructive interference in transport through annulene molecules}

\label{annu}

In this section, we show that the minimal model describing $\pi $ orbitals
for annulenes connected to conducting leads at appropriate places, and for one
electron added or removed from the neutral configuration (so that the ground
state has an odd number of electrons and Kondo effect becomes possible),
corresponds to the model described above, with perfect destructive
interference for $\delta =0$.

The model that describes the $\pi $ orbitals is a Hubbard 
or Pariser-Parr-Pople model \cite{rinc} in a ring of $N=4n$ or $N=4n+2$ sites, with $N\pm 1$ electrons. 
The ground state
is doubly degenerate, with wave vectors $K_{2}=-K_{1}$. We assume that one
of the leads (the left one) is connected to site $j_{L}$. Without loss of
generality, changing the phases of the states 1 and 2 if necessary, one can
take that the couplings of these states to the leads are real and positive
and by symmetry $V_{1}^{L}=V_{2}^{L}>0$. Using  symmetry arguments, for the
right lead connected at site  $j_{R}$, the matrix element of the hybridization 
between both states of the molecule and the right lead are \cite{rinc,benz}

\begin{equation}
V_{i}^{R}=t_{R}\exp \left[ -iK_{i}(j_{R}-j_{L})\right] ,  \label{vr0}
\end{equation}%
where $t_{R}>0$. For symmetric coupling to the leads  $t_{R}=V_{1}^{L}$. 

For $N=4n$,  $K_{i}=\pm \pi /2$. Changing the phase of the conduction
electrons at the right lead one can fix $V_{1}^{R}$ to be real and positive and then
using Eq. (\ref{vr0})

\begin{equation}
V_{2}^{R}=V_{1}^{R}\exp \left[ -i(K_{2}-K_{1})(j_{R}-j_{L})\right]
=(-1)^{j_{R}-j_{L}},  \label{v2r}
\end{equation}%
so that for odd differences between the position of the leads one has
perfect DESINT according to the results of the previous
section (the sum and difference of the states 1 and 2 diagonalize the matrix
$\sum_{ij\nu}\bar{V}_{i}^{\nu }V_{j}^{\nu }d_{j\sigma }^{\dagger }d_{i\sigma }$).

For $N=4n+2$, the last occupied electrons have wave vectors $\pm n\pi /(2n+1)
$ and the first unoccupied ones $\pm (n+1)\pi /(2n+1)$. Therefore we take 
$K_{1}=m\pi /(2n+1)$, with $m=n$ ($n+1$) for one added hole (electron) and
choose the situation with $m$ odd. We propose to connect the right lead at
90 degrees with respect to the left one, so that the former is equally
coupled at the sites $j_{R}=n$ and $j_{R}=n+1$. Then  Eq. (\ref{vr0}) should
be replaced with

\begin{equation}
V_{i}^{R}=2t_{R}\cos (K_{i}/2)\exp \left[ -iK_{i}(n+1/2)\right] .
\label{vi2}
\end{equation}

As before one can change the phases of the right lead so that $V_{1}^{R}$ is
real and positive, what leads to
\begin{eqnarray}
V_{1}^{R} &=&2t_{R}\cos (K_{i}/2),  \label{vra} \\
V_{2}^{R} &=&V_{1}^{R}(-1)^{m},  \nonumber
\end{eqnarray}

so that again there is perfect DESINT for odd $m$. 

It is easy to see that changing the angle between the leads allows for large conductance again,
as in the case of benzene in the \textit{ortho} or \textit{meta} positions \cite{benz}.


\begin{thebibliography}{99}

\bibitem{naturefocus} Aradhya S V and  Venkataraman L,
2013 \textit{Nature Nanotechnology} {\bf 8} 399 

\bibitem{cuevas} Cuevas J C and Scheer E, 2010 \textit{Molecular Electronics:
An Introduction to Theory and Experiment} (World Scientific, Singapore)

\bibitem{carda} Cardamone D, Stafford C, and Mazumdar S, 2006 
\textit{Nano Lett.} \textbf{6} 2422

\bibitem{ke} \textit{Quantum-Interference-Controlled Molecular Electronics},
Ke S-H, Yang W, and Baranger H, 2008
\textit{Nano Lett.} \textbf{8} 3257 

\bibitem{bege} Begemann G, Darau D, Donarini A, and Grifoni M, 2008
\textit{Phys. Rev. B} \textbf{77} 201406(R); \textbf{78} 089901(E) 

\bibitem{dona} Donarini A, Begemann G, and Grifoni M, 2009 
\textit{Nano Lett.} \textbf{9} 2897

\bibitem{rinc} Rinc\'{o}n J, Hallberg K, Aligia A A, and Ramasesha S, 2009 
\textit{Phys. Rev. Lett.} \textbf{103} 266807; references therein.

\bibitem{benz} Tosi L, Roura-Bas P, and Aligia A A, 2012
\textit{J. Phys. Condens. Matter} \textbf{24} 365301; references therein.

\bibitem{garner} Garner M H, Li H, Chen Y, Su T A, Shangguan Z, Paley D W, Liu T, Ng F, Li H, Xiao S, 
Nuckolls C, Venkataraman L, and  Solomon G C, 2018 
\textit{Nature} {\bf 558} 415 

\bibitem{li} Li Y, Buerkle M, Li G, Rostamian A, Wang H, Wang Z, Bowler D R, Miyazaki T, Xiang L, Asai Y, Zhou G,
and Tao N, 2019
\textit{Nat. Mater.} \textbf{18} 357 

\bibitem{bai} Bai J, Daaoub A, Sangtarash S, Li X, Tang Y, Zou Q, Sadeghi H, Liu S, Huang X, Tan Z, 
Liu J, Yang Y, Shi J, M\'esz\'aros G, Chen W, Lambert C, and Hong W, 2019
\textit{Nat. Mater.} \textbf{18} 364  

\bibitem{jan} van Ruitenbeek J M, 2012 
\textit{Physics} \textbf{5} 85

\bibitem{gued} Gu\'edon C M, Valkenier H, Markussen T, Thygesen K S, Hummelen J C, and van der Molen S J, 2012
\textit{Nature Nanotech.} \textbf{7} 305

\bibitem{ball}  Ballmann S, H\"artle R, Coto P B,
Elbing  M, Mayor M, Bryce M R, Thoss M, and Weber H B, 2012  
\textit{Phys. Rev. Lett.} \textbf{109} 056801.

\bibitem{yu3} Yu P, Koci\'{c} N, Repp J, Siegert B, and Donarini A, 2017
\textit{Phys. Rev. Lett.} \textbf{119} 056801; references therin.

\bibitem{hewson} Hewson A C, \textit{The Kondo Problem to Heavy Fermions}
(Cambridge University Press, Cambridge, England, 1997), ISBN 9780521599474.

\bibitem{cro} Cronenwet S M, Oosterkamp T H and Kouwenhoven L P,
1998 \textit{Science} \textbf{281} 540

\bibitem{keller} Keller A J, Amasha S, Weymann I, Moca C P, Rau I G,
 Katine J A, Shtrikman H, Zar\'and G and Goldhaber-Gordon D, 2014 \textit{Nat. Phys.} \textbf{10} 145
 
\bibitem{liang} Liang W, Shores M P, Bockrath M, Long J R, and Park H, 2002 
\textit{Nature} \textbf{417 } 725 

\bibitem{parks2} Parks J J, Champagne A R, Costi T A, Shum W W, Pasupathy A N, 
 Neuscamman E, Flores-Torres S, Cornaglia P S, Aligia A A, Balseiro C A, 
 Chan G K -L, Abru\~{n}a H D and Ralph D C, 2010 \textit{Science} \textbf{328} 1370    

\bibitem{serge} Florens S, Freyn A, Roch N, Wernsdorfer W, Balestro F,
 Roura-Bas P and Aligia A A, 2011 \textit{J. Phys. Condens. Matter} \textbf{23} 243202; references therein.
 
\bibitem{park} Park H, Park J, Lim A K L, Anderson E H, Alivisatos A P and McEuen P L, 2000 \textit{Nature} \textbf{407} 57 

\bibitem{yu2} Yu L H, Keane Z K, Ciszek J W, Cheng L, Stewart M P, Tour J M, and Natelson D, 2004 \textit{Phys. Rev. Lett.} \textbf{93} 266802

\bibitem{fernandez} Fern\'andez-Torrente I, Franke K J, and Pascual J I, 2008 \textit{Phys. Rev. Lett.} \textbf{101} 217203

\bibitem{rak} Rakhmilevitch D, Koryt\'ar R, Bagrets A, Evers F, and Tal O, 2014 \textit{Phys. Rev. Lett.} \textbf{113} 236603

\bibitem{iancu} Iancu V, Schouteden K, Li Z, and Van Haesendonck C, 2016 \textit{Chem. Commun.} \textbf{52} 11359

\bibitem{haer} H\"artle R, Butzin M, Rubio-Pons O, and Thoss M, 2011 \textit{Phys. Rev. Lett.} \textbf{107} 046802

\bibitem{haer2} H\"artle R, Butzin M, and Thoss M, 2013 \textit{Phys. Rev. B} \textbf{87} 085422

\bibitem{desint} Roura-Bas P, Tosi L, Aligia A A, and Hallberg K, 2011 \textit{Phys. Rev. B} \textbf{84} 073406

\bibitem{restor} Tosi L, Roura-Bas P, and Aligia A A, 2015 \textit{J. Phys.: Condens. Matter} \textbf{27} 335601

\bibitem{nishi} Nishikawa Y, Curtin O J, Hewson A C, Crow D J G, and Bauer J, 2016 \textit{Phys. Rev. B} \textbf{93} 235115

\bibitem{fon} Roura-Bas P, Tosi L and Aligia A A, 2013 \textit{Phys. Rev. B} \textbf{87} 195136; references therein.

\bibitem{sate} Roura-Bas P, Tosi L, and Aligia A A, 2016 \textit{Phys. Rev. B} \textbf{93} 115139

\bibitem{corna1} Cornaglia P S, Ness H, and Grempel D R, 2004 \textit{Phys. Rev. Lett.} \textbf{93} 147201

\bibitem{paaske} Paaske J and Flensberg K, 2005 \textit{Phys. Rev. Lett.} \textbf{94} 176801 

\bibitem{hm} Hewson A C and Meyer D, 2002 \textit{J. Phys.: Condens. Matter} \textbf{14} 427

\bibitem{lili} Arrachea L and Rozenberg M J, 2005 \textit{Phys. Rev. B} \textbf{72} 041301(R)

\bibitem{corna2} Cornaglia P S, Usaj G, and Balseiro C A, 2007 \textit{Phys. Rev. B} \textbf{76} 241403(R)

\bibitem{zitko} \v{Z}itko R and Pruschke Th, 2009 \textit{Phys. Rev. B} \textbf{79} 085106

\bibitem{mon} Monreal R C and Martin-Rodero A, 2009 \textit{Phys. Rev. B} \textbf{79} 115140

\bibitem{yang} Yang K H,  Wu Y P, and Zhao Y L, 2010 \textit{Europhys. Lett.} \textbf{89} 37008 

\bibitem{win} Wingreen N S and Meir Y, 1994 \textit{Phys. Rev. B} \textbf{49} 11040

\bibitem{nca2} Hettler M H, Kroha J and Hershfield S, 1998 \textit{Phys. Rev. B}  \textbf{58} 5649

\bibitem{scali} Roura-Bas P, 2010 \textit{Phys. Rev. B} \textbf{81} 155327

\bibitem{tetta} Tettamanzi G C, Verduijn J, Lansbergen G P, Blaauboer M, Calder\'{o}n M J, Aguado R, and  Rogge S, 2012 \textit{Phys. Rev. Lett.} \textbf{108} 046803

\bibitem{ogu} Oguri A, 2005 {\it J. Phys. Soc. Jpn.} {\bf 74} 110; references therein.

\bibitem{ct} Aligia A A, 2018 \textit{J. Phys. Condens. Matter} \textbf{30} 155304 (2018); references therein.

\bibitem{tk} P\'erez Daroca D, Roura-Bas P, and Aligia A A, 2018 \textit{Phys. Rev. B} \textbf{98} 245406

\bibitem{lim} Lim J S, Choi M -S, Choi M Y, L\'opez R, and Aguado R, 2006 \textit{Phys. Rev. B} \textbf{74} 205119

\bibitem{su42} Tosi L, Roura-Bas P, and Aligia A A, 2012 \textit{Physica B} \textbf{407} 3259

\bibitem{anders} Anders F B, Logan D E, Galpin M R, and Finkelstein G, 2008 \textit{Phys. Rev. Lett.} \textbf{100} 086809

\bibitem{note} For these parameters $\lambda=\sqrt{10}$ and $\Delta=0.05$, the non-equilibrium 
calculations turned out to be very cumbersome. Then we have calculated only equilibrium properties.
They suffice to show the effects on the EVI on $T_K$.

\bibitem{khe} Khedri A, Costi T A, and Meden V, 2017 \textit{Phys. Rev. B} \textbf{96} 195155

\bibitem{meir} Meir Y and Wingreen N S, 1992 \textit{Phys. Rev. Lett.} \textbf{68} 2512

\bibitem{capac} Fern\'andez J, Lisandrini F, Roura-Bas P, Gazza C, and Aligia A A, 2018 \textit{
Phys. Rev. B} {\bf 97} 045144; references therein.

\bibitem{yoshi} Yoshimori A and Zawadowski A, 1982 \textit{J. Phys. C} \textbf{15} 5241 

\bibitem{lang} Langreth D C, 1966 \textit{Phys. Rev.} \textbf{150} 516

\bibitem{datta} Datta S, \textit{Electronic transport in mesoscopic systems}
(Cambridge University Press, Cambridge, England, 2003).

\bibitem{pust} Pustilnik M and Glazman L I, 2001 \textit{Phys. Rev. Lett.} \textbf{87} 216601



\end{thebibliography}
\end{document}